# Multifractal analysis of sentence lengths in English literary texts

Iwona Grabska-Gradzińska [a], Andrzej Kulig [b], Jarosław Kwapień [b], Paweł Oświęcimka [b], Stanisław Drożdż [b,c]

[a] Faculty of Physics, Astronomy and Applied Computer Science, Jagiellonian University,
ul. Reymonta 4, 30-059 Kraków, Poland
[b] Institute of Nuclear Physics, Polish Academy of Sciences, ul. Radzikowskiego 152, 31-342 Kraków, Poland
[c] Faculty of Physics, Mathematics and Computer Science, Cracow University of Technology,
ul. Warszawska 24, 31-155 Kraków, Poland

**Abstract**

This paper presents analysis of 30 literary texts written in English by different authors. For each text, there were created time series representing length of sentences in words and analyzed its fractal properties using two methods of multifractal analysis: MFDFA and WTMM. Both methods showed that there are texts which can be considered multifractal in this representation but a majority of texts are not multifractal or even not fractal at all. Out of 30 books, only a few have so-correlated lengths of consecutive sentences that the analyzed signals can be interpreted as real multifractals. An interesting direction for future investigations would be identifying what are the specific features which cause certain texts to be multifractal and other to be monofractal or even not fractal at all.

## 1. Introduction

Natural language is a highly complicated result of human evolution on both the biological and the social level. According to a recent hypothesis, it developed into a self-organized structure such that human brain can easily and spontaneously learn it during a few years of early childhood [1]. This means that, to a significant degree, the structure of natural language reflects the brain's inner organization. This connection is among the issues that make studying natural language especially interesting even outside the field of linguistics.

Indeed, over recent years the natural language has drawn attention of the researchers whose primary field of interest is information science, physics, and science of complex systems [2]. They attempt to explain certain fundamental properties of language like the Zipf law [3, 4, 5], which can be viewed as scale invariance [6, 7, 8], or to identify other properties which are typical for complex systems, like hierarchical structure [9] or long-range correlations [10]. Especially the latter seems to be interesting as reflecting the principles of brain function which is known to be long-range temporally correlated at least in its certain aspects [11,12,13,14]. As regards the correlation structure of natural language, Ebeling and Pöschel [10] showed that pairs of letters can be correlated over distances of a few hundred letters. Ebeling and Neiman [16] refined this analysis and by using as the Hölder exponents and the Fourier power spectra, they proved that such correlations extend over paragraphs and chapters. Hřebiček [17] considered the variability of sentence lengths and found that their Hurst exponent H≈0.6 which means they are positively linearly correlated over long distances. The Hurst R/S analysis was also applied by Montemurro and Pury [18] to time series of words mapped onto numbers according to the words' frequency ranks. As a result, a convincing evidence was given that word usage by the authors of literary works is long-range correlated even over the corpora consisting of many texts of the same author joined together. This means that the correlations of this type are independent of a given text's information content and are due to an author's style of writing or way of thinking. This result, presented in Ref. [18] only for English texts, was later confirmed for other languages with different grammar and semantics by Bhan et al. [21].

Differences in correlation structure between languages were reported by Şahin et al. [20] who unambiguously mapped letters onto numbers, summed up these numbers for each word separately, and the so-preprocessed texts were then a subject to the detrended fluctuation analysis (DFA) [22], which revealed various scaling regimes in the fluctuations that were different for different languages. Melnyk et al. [19] reduced various text samples to coarse-grained symbolic sequences consisting of 0's (the letters a-m) and 1's (n-z) and communicated that even in this case the language exhibits correlations that can be either negative for short, intra-sequence ranges, or positive for long ranges.

Most of the above studies were largely restricted to linear correlations because of the limitations of the applied methods. However, the scaling properties of different representations of literary texts and their hierarchical organization suggest that they can possess fractal structure.

Going a step further one can ask whether this structure is completely homogeneous or rather it comprises nonhomogeneous, multifractal features. Pavlov et al. [23] mapped a text into a point process defined by the intervals between the successive occurrences of specific combination of letters and found that the corresponding sequence is multifractal and presents nonlinear long-range correlations. Similar results for the word-frequency and the word-length time series constructed for L. Carroll texts were communicated by Gillet and Ausloos [15]. An interesting feature of that study is comparison of a natural (English) and an artificial (Esperanto) language which have different multifractal properties.

## 2. Methods

Here we consider a different type of text representation: the sentence lengths as measured by the number of words. We choose this particular representation because due to the fact that single sentence often comprises a well-distinguished piece of information, it may somehow reflect the process of thinking. Technically, each sentence in a text is identified by the standard punctuation marks: full-stop, colon, semicolon, interrogation mark and exclamation mark. We neglect commas as in many circumstances they do not distinguish minimum pieces of information (for instance, when they are auxiliary used to separate listed elements or to avoid ambiguity of a message). We count words that appear between consecutive sentence-closing marks and form a time series consisting of the corresponding numbers in a preserved order. We investigate possible nonlinear statistical dependences in such data by considering fractal properties of its structure.

Our principal method of numerical study is the multifractal detrended fluctuation analysis (MFDFA) [24]. We also apply the wavelet transform modulus maxima (WTMM) method [25] as an auxiliary tool which can make the results of MFDFA more trustful (the use of WTMM as a basic tool is not recommended due to its lesser reliability for short signals [26]).

*2.1. MFDFA*

Let assume that we have a time series of numbers $x(i)$ where $i = 1, 2, \ldots, N$ denotes the consecutive sentences. For this time series, one needs to estimate the signal profile [24]:

$$Y(i) = \sum_{k=1}^{i}(x(i) - \langle x \rangle), \qquad (1)$$

where $\langle \ldots \rangle$ denotes the mean of $x(i)$ taken over the whole series. $Y(i)$ can now be divided into $M$ disjoint segments of length $n$ starting from the beginning of the time series $\{x\}$. For each segment $v$, $v = 1, \ldots, M$, one calculates a local trend by least-squares fitting the polynomial $P_v^{(l)}$ of order $l$ to the signal segment. Then the variance:

$$F^2(v, n) = \frac{1}{n} \sum_{j=1}^{n} \{Y[(v-1)n + j] - P_v^{(l)}(j)\}^2. \qquad (2)$$

has to be derived. In order to avoid neglecting the data points at the end of $\{x\}$ that do not fall into any segment, the same procedure is repeated for $M$ segments starting from the end of the signal. In result, one obtains $2M$ segments total and the same number of values of $F^2$. The polynomial order $l$ can be equal to 1 (DFA1), 2 (DFA2). Finally, the variances (2) have to be averaged over all the segments $v$, which leads to the $qth$ order fluctuation function:

$$F_q(n) = \left\{\frac{1}{2M} \sum_{v=1}^{2M} [F^2(v, n)]^{q/2}\right\}^{1/q}, q \in R \qquad (3)$$

The key step is now to determine the statistical dependence of $F_q$ on $n$, which can be done after calculating $F_q(n)$ for many different segment lengths $n$. The rationale behind this procedure is that if the analysed time series has fractal properties, the fluctuation function reveals the power-law scaling $F_q(n) \sim n^{h(q)}$ (4) for large $n$. The family of the scaling exponents $h(q)$ can be obtained in this way by using different values of $q$. The exponents $h(q)$ can be considered a generalization of the Hurst exponent $H$ with the special case of $H = h(2)$. Multifractals can be distinguished from monofractals by looking at $h(q)$: if $h(q) = H$ for all $q$, then the signal under study is monofractal; it is multifractal otherwise.

From $h(q)$, one can calculate the Hölder exponents $\alpha$ and the singularity spectrum $f(\alpha)$ using the following relations (e.g. [22]):
$\alpha = h(q) + qh'(q), \; f(\alpha) = q[\alpha - h(q)] + 1$ (5)   where $h'(q)$ denotes the derivative of $h(q)$ with respect to $q$.

*2.2 WTMM*

WTMM method exploits the existence of scaling properties of wavelet transform coefficients for fractal signals [23]. The wavelet transform is defined by the following relation:

$$T_\psi(n, s) = \frac{1}{s} \sum_{i=1}^{N} \psi\left(\frac{i-n}{s}\right) x(i) \qquad (6)$$

where $\psi$ is a wavelet kernel shifted by $n$ and $s$ is scale. It decomposes a signal in time-scale plane. In principle, a mother wavelet $\psi$ can be chosen arbitrarily, but in practice it should well reproduce the features of a studied signal. The family of wavelets which is used most frequently in the context of time series is the $mth$ derivative of a Gaussian:

$$\psi^{(m)}(x) = \frac{d^m}{dx^m}\left(e^{\frac{-x^2}{2}}\right) \qquad (7)$$

working well in removing the signal trends approximated by polynomials up to $(m-1)th$ order [25].

A singularity present in data leads to a power-law behaviour of the coefficients $T_\psi$:
$$T_\psi(n_0, s) \sim s^{\alpha(n_0)}. \quad (8)$$
Since this relation might be not stable in the case of densely packed singularities, it is suggested to identify the local maxima of $T_\psi$ and then calculate the partition function from their moduli:
$$Z(q,s) = \sum_{l \in L(s)} |T_\psi(n_l(s), s)|^q. \quad (9)$$
Here, $L(s)$ is the set of all maxima for scale $s$ and $n_l(s)$ is the position of a particular maximum. Monotonicity of $Z(q,s')$ on $s'$ can be preserved by adding a supremum condition:
$$Z(q,s) = \sum_{l \in L(s)} \left( \sup_{s' \leq s} |T_\psi(n_l(s'), s')| \right)^q. \quad (10)$$
For a fractal signal, $Z(q,s) \sim s^{\tau(q)}$.

The singularity spectrum $f(\alpha)$ can be calculated according to the following formulas [27]:
$$\alpha = \tau'(q) \text{ and } f(\alpha) = q\alpha - \tau(q). \quad (11)$$
Similar to the above $h(q)$ functions, if $\tau'(q)$ is linear, it indicates a monofractal signal, while its nonlinear behaviour suggests a multifractal one.

## 3. Results

We apply the both above methods to time series representing the sentence lengths of 30 randomly selected English literary texts taken from the Gutenberg Project page (5 books by C. Dickens, 4 by J. Austen, 3 books by each of J. Joyce, A. Conan Doyle, and M. Twain, and 1 book by each of O. Wilde, A. Christie, H. Melville, L. Carroll, E.R. Borroughs, C. Darwin, U. Sinclair, J. Swift, M. Shelley, and B. Stoker). In order to obtain statistically significant results, each text consists of at least $N \approx 2 \cdot 10^3$ sentences ("Adventures of Alice in Wonderland" by L. Carroll) with the longest signals reaching $N \approx 25 \cdot 10^3$ ("Ulysses" by J. Joyce and "Bleak House" by C. Dickens).

Interpretation of the fluctuation function behaviour is a delicate matter [28]. The family $F_q(n)$ can be considered representing multifractal data without any doubt only when the range of n for which $F_q(n)$ is power-law extends over almost whole possible values of n. For real signals, however, this criterion is typically not met and a scaling range is much shorter. A typical case in this respect is such that the multifractal scaling of $F_q(n)$ is seen for some $n < n_0$ and there is only monofractal scaling for $n > n_0$ with $n_{min} < n_0 < n_{max}$. If this is the case, the interpretation of scaling has to be done with care, based on the results of model data with known fractal properties and one's own experience [28]. There are two possible interpretations of such result depending on $n_0$. First, if $n_0 \ll N$ and surrogate data (for example, consisting of randomized original signals) show also a trace of multifractal scaling below the even smaller threshold $n_0$, it means that the data under study is in fact monofractal (a single point in a graph of $f(\alpha)$) or bifractal (two points) but highly nonstationary, and this nonstationarity together with possible "fat tails" of the corresponding pdf give the apparent multifractal behavior of $F_q(n)$. Second, if the range of scaling is long enough (more than one decade long) and $n_0$ is a significant fraction of N, as well as the surrogate data produces a substantially less multifractal behavior of $F_q(n)$. (i.e., $h(q)$ is much less nonlinear and the $f(\alpha)$ parabola is narrower) than in the case of the original signals, one may infer that the analysed data is indeed multifractal. Sometimes, the multifractal character of data is accompanied by a long power-law relaxation of the autocorrelation function, but this connection is not always observed.

Fig. 1 shows examples of the fluctuation function $F_q(n)$ (Eq. (3)) for four texts with different fractal properties: a text without any clear fractal structure (no scaling range of $F_q(n)$, (a)), a text with an evident monofractal structure (b), a text with rather spurious multifractal-like structure for small scales n (c), and a text which can be considered real multifractal (d). Each of the 30 texts considered in our study can be assigned to one of these classes. Fig. 2 presents the family of fluctuation functions calculated for real texts (the same as in Figure 1(c) and 1(d)) together with their counterparts for the respective randomized signals. Fig. 3 shows the singularity spectra $f(\alpha)$ for those texts for which this was possible. A comparison of the results obtained with MFDFA and WTMM for an exemplary text is shown in Fig. 4. It should be noted that the more convincing is the multifractality of the data, the closer results are obtained by means of the two methods. Finally, Fig. 5 shows the autocorrelation function for the same four texts as in Fig. 1. As one can clearly see, only in the last example (Fig. 3(d)), the function is power law for some range of n. This confirms our conclusion about multifractality of the underlying text.

Out of 30 books, only a few have so-correlated lengths of consecutive sentences that the analysed signals can be interpreted as real multifractals. Although we observe that for some authors (Twain, Conan Doyle) the calculated fractal properties are roughly invariant under a change of texts. For others, different texts can have different properties (Austen). An interesting direction for future investigations would be identifying what are the specific features that cause certain texts to be multifractal and other to be monofractal or even not fractal at all.

## 4. Conclusions

We analysed 30 literary texts written in English by 17 different authors. For each text, we created a time series representing length of sentences in words and analysed its fractal properties using two methods of multifractal analysis: MFDFA and WTMM. Both methods showed that there are texts which can be considered multifractal in this representation, but a majority of texts are not multifractal. We obtained no indication what is the origin of multifractality in natural language samples, but there is at least some evidence that, in some cases, the fractal properties can be similar for different works of the same author. Our results suggest that the fractal properties of natural language are a subtle problem which requires much more study in future.

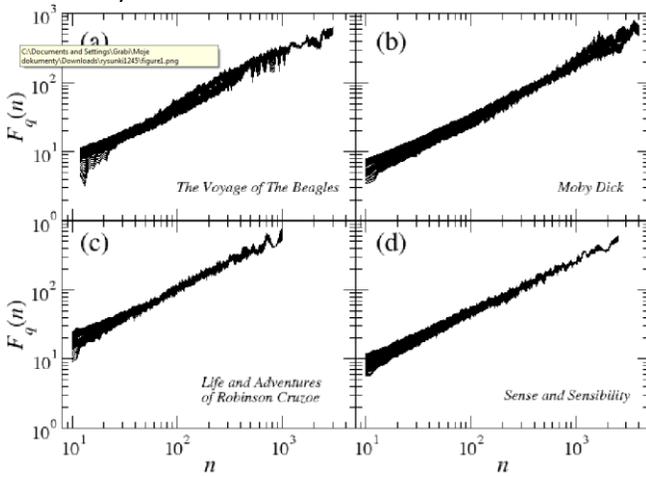

FIGURE 1: Fluctuation function $F_q(n)$ calculated for time series of sentence lengths (in words) for 4 representative texts. Since such time series can be leptokurtic, the Rényi parameter was restricted to −3≤q≤3. While no fractal structure can be observed in (a), plot in (d) is convincingly multifractal.

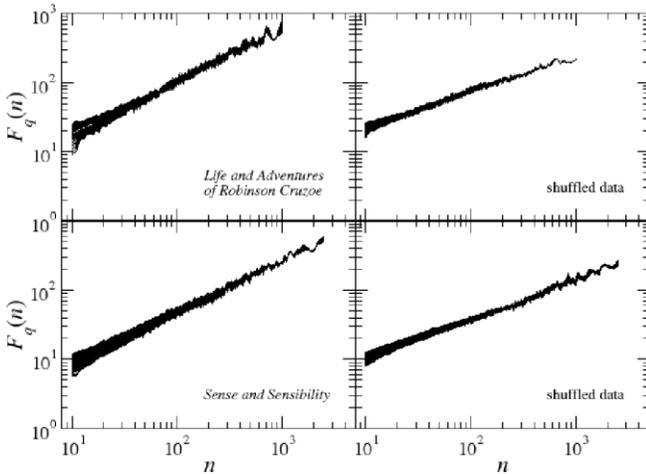

FIGURE 2: Comparison of fluctuation functions $F_q(n)$ for real (left column) and shuffled data (right column) for a text with rather spurious multifractality and a text with real multifractality.

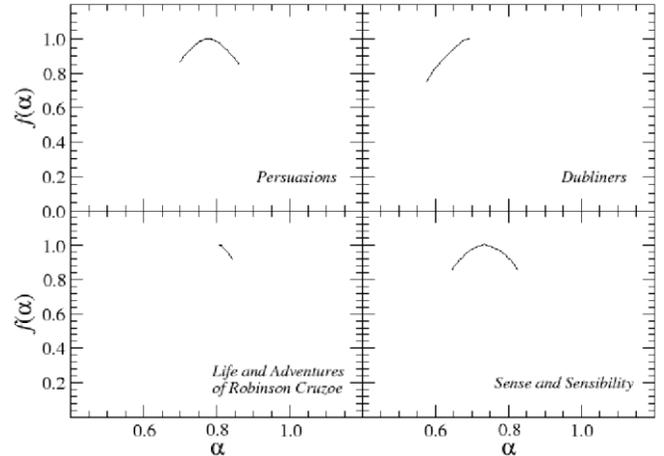

FIGURE 3: Singularity spectra $f(\alpha)$ for different texts. The broader the spectrum, the richer is the fractal structure of data. The texts for which the width of $f(\alpha)$ substantially exceeds 0.1 can be considered multifractal (a criterion based on the authors' own experience).

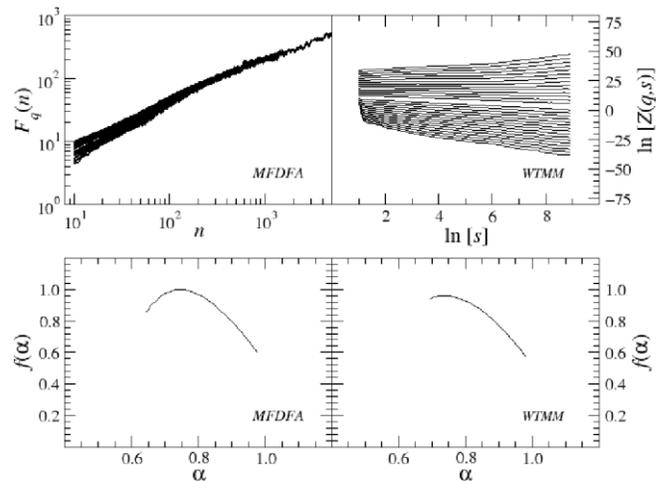

FIGURE 4: Comparison of results obtained for the same text of "David Copperfield" by C. Dickens with two different methods of multifractal analysis: MFDFA (left) and WTMM (right).

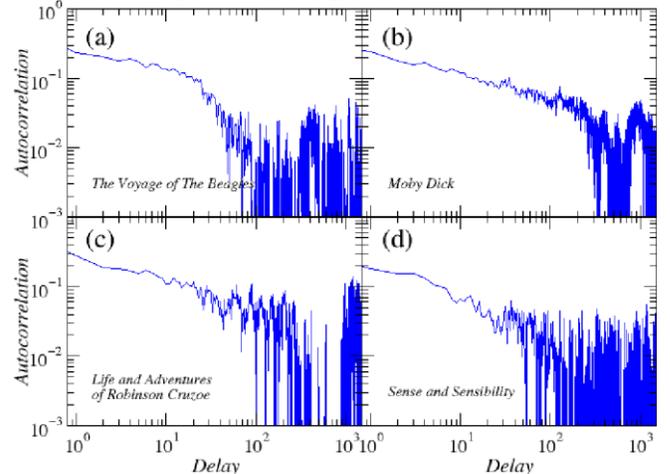

FIGURE 5: Autocorrelation function for the same texts as in Fig. 1 plotted in log-log scale.